\documentclass[acus]{JAC2000}

\usepackage{graphicx}

\begin{document}
\title{\flushright{W03}\\[15pt] \centering 
QCD Studies in Two-Photon Collisions at CLEO}
\author{Vladimir Savinov, University of Pittsburgh, Pittsburgh, PA 15260, USA}

\maketitle

\begin{abstract}

We review the results of two-photon measurements 
performed up to date by the CLEO experiment 
at Cornell University, Ithaca, NY. 
These measurements provide an almost background-free 
virtual laboratory to study strong interactions 
in the process of the $e^+e^-$ scattering. 
We discuss the measurements of 
two-photon partial widths of charmonium,  
cross sections for hadron pairs production, 
antisearch for glueballs and 
the measurements of $\gamma^*\gamma \to$ pseudoscalar meson 
transition form factors. 
We emphasize the importance of other possible analyses, 
favorable trigger conditions and selection criteria 
of the presently running experiment 
and the advantages of CLEOc 
-- the future $\tau$-charm factory 
with the existing CLEO III detector. 

\end{abstract}

\section{Introduction}

One of the ways to study properties of strong interactions, 
surprisingly, is to collide high energy photons. 
Photons do not interact strongly, however, in presence of 
other photons they can fluctuate into quark pairs that 
have sizable probability to realize as hadrons. 
Space-like photons of relatively high energies can be obtained 
in the process of the $e^+e^-$ scattering, {\it i.e.} in the 
$e^+e^- \to e^+e^-$hadrons reactions, where hadrons are 
produced in charge-even, {\it i.e.} $C=+1$ state. 
These processes proceed mainly by two-photon fusion 
therefore telling us about the strength of relevant 
$\gamma\gamma$ couplings 
and the properties of particles born in such reactions. 
When a single hadron is born, the production cross section 
is proportional to its two-photon partial width 
thus allowing the measurement of this quantity in the 
time-reversed two-photon decay. 
When (at least) one of the photons is substantially off-mass shell, 
we can measure the form factors associated with two-photon 
transitions that probe spatial distribution of electric charge inside of 
produced hadrons thus telling us about respective wave functions 
{\it i.e.} details of binding potential. 
The kinematics of two-photon collisions in the 
$e^+e^-$ scattering is described elsewhere\cite{budnev,poppe}. 

\section{CLEO Experiment}

The results discussed in this short review 
were obtained from 
the data collected at the 
Cornell Electron Storage Ring (CESR) 
with the CLEO detector. 
The results are based on statistics that correspond 
to an integrated $e^+e^-$ luminosity of 
up to $9.2 fb^{-1}$ collected at the $\Upsilon(4{\rm S})$ energy 
of 10.58 GeV and up to $4.6 fb^{-1}$ 
collected approximately 60 MeV below the $\Upsilon(4{\rm S})$ energy. 
Our data sample was recorded with two configurations 
of the CLEO detector. 
The first third of the data were recorded with 
the CLEO II detector\cite{CLEOII_description} 
which consisted of three cylindrical drift chambers 
placed in an axial solenoidal magnetic field of 1.5T, 
a CsI(Tl)-crystal electromagnetic calorimeter, 
a time-of-flight (TOF) plastic scintillator system 
and a muon system (proportional counters embedded 
at various depths in the steel absorber). 
Two thirds of the data were taken with 
the CLEO II.V configuration of the detector 
where the innermost drift chamber 
was replaced by a silicon vertex detector\cite{CLEOII.V_description} 
and the argon-ethane gas of the main drift chamber 
was changed to a helium-propane mixture. 
This upgrade led to improved resolutions in momentum 
and specific ionization energy loss 
($dE/dx$) measurements. 
\newline \indent The three-tier CLEO trigger system\cite{CLEOII_trigger} complemented by the 
software filter for beam-gas rejection utilizes 
the information from the two outer drift chambers, 
the TOF system and electromagnetic calorimeter. 
The response of the detector is modeled with 
a GEANT-based\cite{GEANT} Monte Carlo (MC) simulation program. 
The data and simulated samples are processed 
by the same event reconstruction program. 
Whenever possible the efficiencies are either 
calibrated or corrected for the difference 
between simulated and actual detector responses 
using direct measurements from independent data. 
This is especially important for understanding the trigger 
efficiency as most two-photon events experience strong Lorentz 
boost along the $e^+e^-$ collision axis 
often missing detection and failing to trigger data taking. 
\newline \indent All analyses presented in 
this summary employ complete reconstruction of hadronic final 
states born in the process of two-photon fusion. 
In all but the form factor measurements, final state leptons 
escape detection in the beam pipe because of kinematics of two-photon 
collisions that favors small scattering angles for electron and 
positron. The detailed descriptions of the reviewed CLEO analyses 
can be found in the references 
to CLEO papers provided in the bibliography section. 
Relevant theoretical references can be found in 
the CLEO papers. 

\section{Charmonium Measurements}

CLEO measured two-photon partial widths 
of $\chi_{c_2}$ in the $J/\psi \gamma$ final state\cite{cleo_charm_old}, 
and, more recently, of the $\eta_c$ in the $K_s K^{\pm} \pi^{\mp}$ 
final state\cite{cleo_etac} 
and $\chi_{c_0}$ and $\chi_{c_2}$ in the $\pi^+\pi^-\pi^+\pi^-$ 
decays\cite{cleo_charm_new}. 
The most recent results are 
$\Gamma_{\gamma\gamma}(\chi_{c_0}) = (3.76 \pm 0.65 ({\rm stat}) \pm 0.41 ({\rm syst}) \pm 1.69 ({\rm br}))$ keV, 
$\Gamma_{\gamma\gamma}(\chi_{c_2}) = (0.53 \pm 0.15 ({\rm stat}) \pm 0.06 ({\rm syst}) \pm 0.22 ({\rm br}))$ keV 
and 
$\Gamma_{\gamma\gamma}(\eta_c) = (7.6 \pm 0.8 ({\rm stat}) \pm 0.4 ({\rm syst}) \pm 2.3 ({\rm br}))$ keV. 

Our results on two-photon partial widths of charmonium 
are consistent with some of the theoretical predictions 
we refer to in our publications. It should be emphasized 
that the extraction of $\alpha_s$ from our data presented 
in our papers was done mainly 
to compare our results with other similar measurements. 
As became known recently\cite{kirill}, 
theoretical attempts to include next-to-next 
to leading order corrections in $\alpha_s$ 
to perturbative Quantum Chromodynamics (pQCD) predictions 
for two-photon partial widths 
diverge and fail thus making 
such $\alpha_s$ extraction poorly defined. 
Another important aspect of the analyses presented 
in our papers on charmonium is the assumption 
about absence of the interference between 
resonant and continuum two-photon production 
of the studied final states. In the new $\eta_c$ 
analysis where we had sufficient statistics 
to study possible effect of such interference, 
we found no convincing indication of this effect. 
Therefore, no interference was taken into account 
when estimating systematic effects in either of our 
charmonium analyses. 

Our new result for the $\chi_{c_0}$ is consistent 
with previous result\cite{cleo_charm_old} obtained in the $J/\psi \gamma$ mode. 
Also, our measurement of the product of $\eta_c$ two-photon partial width 
and $\eta_c \to K_s K \pi$ branching fraction 
is consistent with our preliminary results\cite{cleo_etac_preliminary}. 
We would like to alert the reader to the fact that 
there is a large uncertainty in our measurements 
of two-photon partial widths as we measure the 
products of these with the branching fractions 
for the final states where we reconstruct 
charmonium. Therefore we inherit large uncertainties in the 
experimental values for these branching fractions 
when we convert the measured products to the 
measurements of two-photon partial widths. 
Great care should be executed when comparing 
the results of different experiments as a more recent 
experiment often uses an updated value for the final state 
branching fraction as an older one. A good strategy would be 
to have old editions of the review of particle properties 
available for such comparisons. 

In our $\eta_c$ analysis we also measured the mass and 
(total) width of this charmonium state: 
$M(\eta_c) = (2980.4 \pm 2.3 ({\rm stat}) \pm 0.6 ({\rm syst}))$ MeV 
and 
$M(\eta_c) = (27.0 \pm 5.8 ({\rm stat}) \pm 1.4 ({\rm syst}))$ MeV. 
While we did a thorough study of systematics 
that could be a source of experimental error, 
we have to emphasize that we have no calibration channel 
that would be a ``golden-bullet'' kind of a {\it proof} that 
we understand the mass and width measurements around 3 GeV 
in the four charged tracks final states 
at CLEO. 
This is to provide the reader with more information, 
not to give an impression that we have any doubts in our results. 
We refer the reader to our publication on the subject\cite{cleo_etac} 
for more information. 

\section{Hadron Pair Production}

\begin{figure*}[t]
\centering
\centerline{
\includegraphics*[width=100mm,height=110mm]{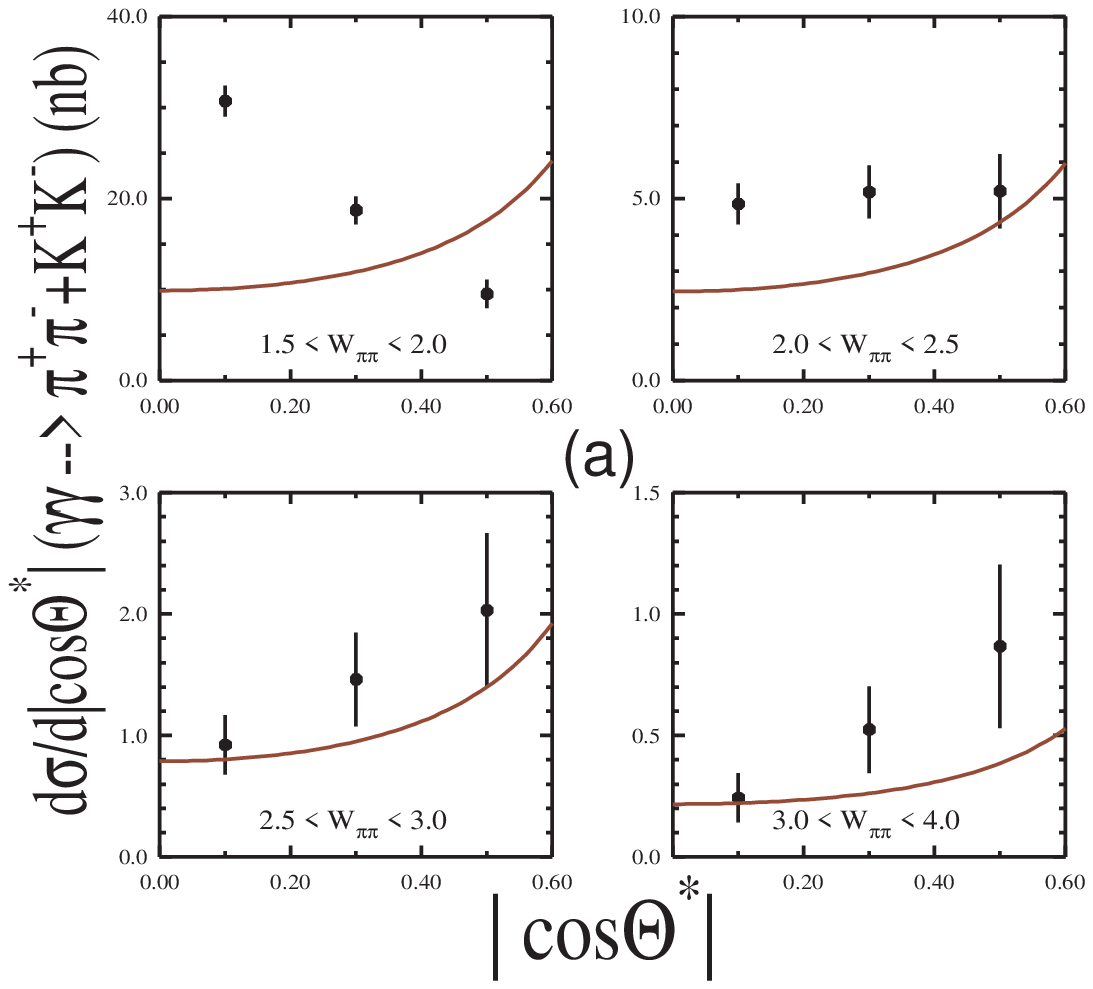}
\includegraphics*[width=90mm,height=110mm]{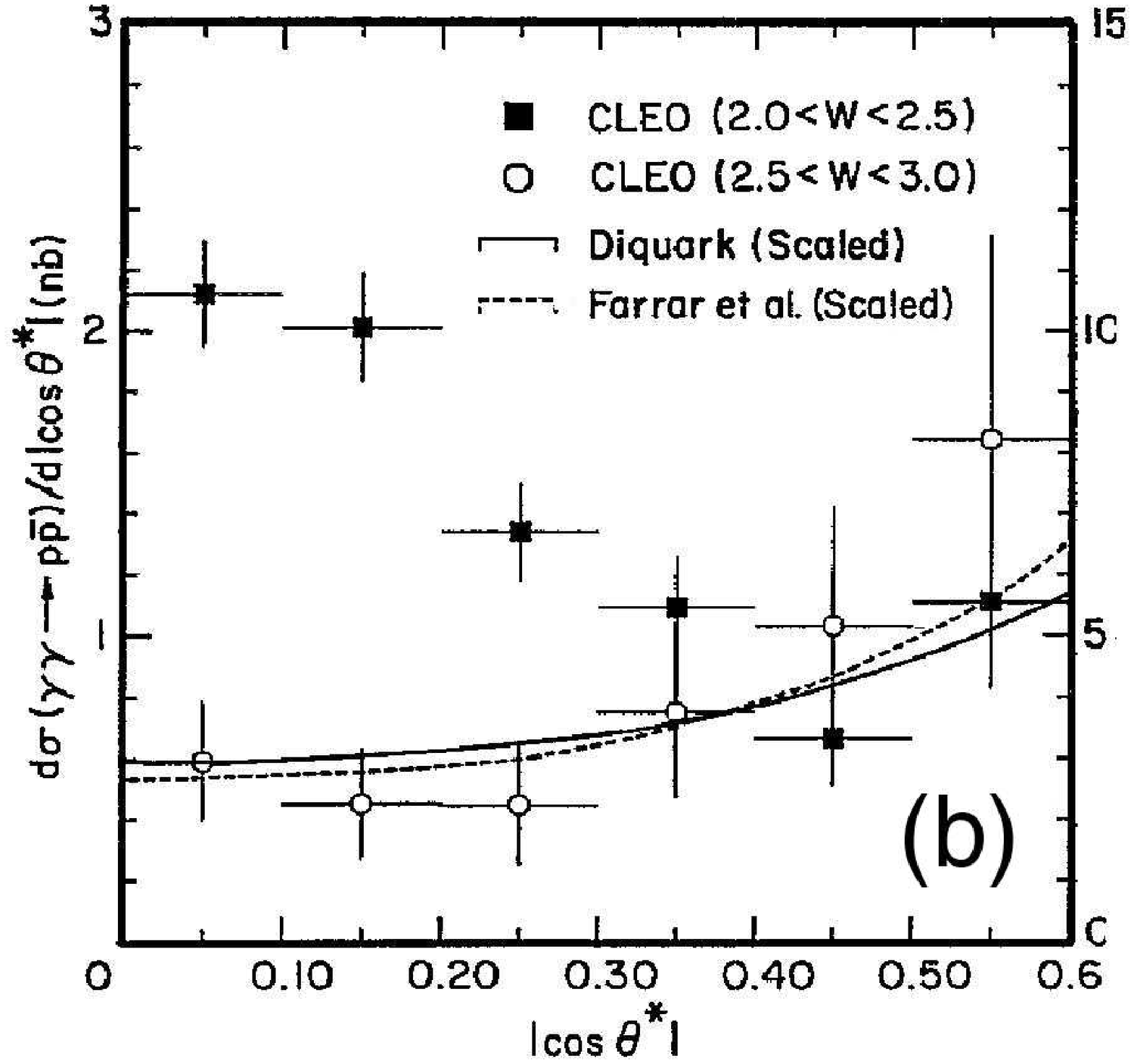}
}
\caption{CLEO results on (a) $\pi\pi$ and (b) $p\bar{p}$ pairs production. 
See the text for more information.}
\label{fig_cleo_pairs}
\end{figure*}

CLEO measured a number of cross sections for two-photon 
production of hadron pairs. These include combined measurement for 
$\pi^+\pi^-$ and $K^+K^-$ pairs\cite{cleo_pipi}, 
$p\bar{p}$ pairs\cite{cleo_pp} and $\Lambda\bar{\Lambda}$ pairs\cite{cleo_lamlam}. 
Our results agree well with the predictions of perturbative QCD and 
diquark model, especially at higher invariant masses of produced pairs. 
The agreement for the values and shapes of the cross sections 
is also reasonable in the region of relatively low pair masses 
and this fact is quite surprising because 
predictions based on perturbative QCD are not expected to hold there. 
However, and more important, our result prove that there is a qualitative difference 
between hadron pairs produced at lower masses and at higher masses where 
the definitions of lower and higher are CLEO-specific and are determined by 
energies available to us in our experiment. This qualitative difference 
is demonstrated in Fig.\ref{fig_cleo_pairs}(a) and Fig.~\ref{fig_cleo_pairs}(b) for 
$\gamma\gamma \to \pi^+\pi^-$ and $\gamma\gamma \to pp$ measurements, respectively. 
These figures show efficiency-corrected and background-subtracted 
distributions of our data (points with the error bars) 
for several intervals of hadron pairs invariant mass 
versus $|\cos{\theta^*}|$,  where $\theta^*$
is helicity angle. 
Curves in figures show (a) perturbative prediction\cite{BL} 
made by Brodsky and Lepage assuming their 
mechanism $\gamma\gamma \to q\bar{q}g \to \pi^+\pi^-$ (BL) 
for pion pairs production and (b) diquark model\cite{kroll_diquark} 
and perturbative\cite{farrar} predictions 
for $p\bar{p}$ production. 
Theoretical curves 
shown in Fig.~\ref{fig_cleo_pairs}(b) 
are normalized to our data and are displayed 
only for $p\bar{p}$ invariant mass above 2.5 GeV. 
Notice that there are two vertical scales for 
two distributions shown in Fig.~\ref{fig_cleo_pairs}(b), 
the right-side scale is for the events collected at 
higher invariant mass. 
Helicity angle is measured 
between the direction of one of the hadrons 
in the rest frame of two photons 
and the momentum direction for a pair in the laboratory 
reference frame. Notice that the range of helicity angle-related 
variable is restricted to be below 0.6 which is a typical acceptance 
region for a two-photon experiment. 
It would take photons to be highly off-mass shell 
to extend the range of non-zero acceptance for cosine of helicity angle. 
This analysis is being planned, meanwhile, notice that 
the distribution for the hadron pairs in the region 
of higher invariant mass shows clear transition to the 
diffractive ({\it i.e.} perturbative) behavior when compared to 
that for the pairs in the region of lower invariant mass. 

\section{Glueball Antisearch}

\begin{figure*}[t]
\centering
\includegraphics*[width=180mm,height=90mm]{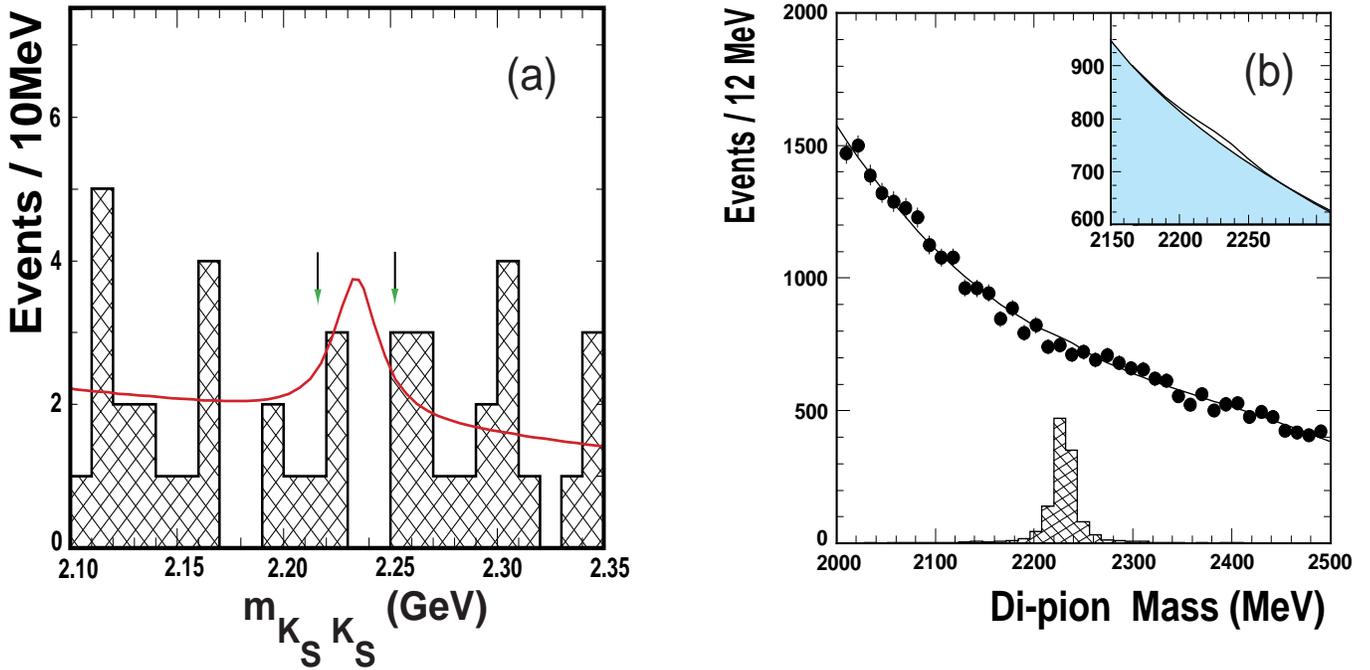}
\caption{CLEO antisearch for $f_j(2220)$ in (a) $K_s K_s$ and (b) $\pi^+\pi^-$ channels. 
See the text for more information.}
\label{fig_cleo_glue}
\end{figure*}

Possible existence of glueballs, {\it i.e.} hadrons 
made of constituent glue, does not 
contradict to known experimental and theoretical facts. 
More important, such states are predicted to exist by 
calculations on the lattice (LQCD). The main caveat here is that 
these predictions are made in the so-called quenched approximation, 
when quenching is removed, there is no consensus yet 
if the predictions are going to hold. Therefore, possible discovery of 
glueballs should help to advance the theory. 
On the other hand, it is also possible that no 
glue bound states could ever exist and this 
scenario would not be a great disappointment, neither 
a catastrophe for LQCD. 
If the latter non-existence scenario realizes in nature, 
it is still possible that glueball-like 
field configurations play an important role in non-perturbative 
QCD processes acting as a mass scale that modifies 
predictions for cross sections at relatively low  
energies, {\it i.e.} below 10 GeV. 
\newline \indent So far CLEO has only searched for the most famous glueball candidate, 
$f_j(2220)$ observed a few years ago in radiative $J/\psi$ decays at BES. 
We searched for this resonance in the $K_s K_s$ and $\pi\pi$ final 
states and set 95\% CL upper limits on the products of its two-photon partial 
width and relevant branching fractions of 
$\Gamma_{\gamma\gamma}(f_j(2220)) {\cal B}(f_j(2220) \to K_s K_s) < 1.3$eV 
and 
$\Gamma_{\gamma\gamma}(f_j(2220)) {\cal B}(f_j(2220) \to \pi^+\pi^-\pi^+\pi^-) < 2.5$eV, 
respectively. 
Small values of these upper limits are not surprising 
as if the $f_j(2220)$ state existed, its electromagnetic coupling 
would be very small because gluons do not couple 
to photons directly. Invariant mass plots for relevant final states 
from our analyses 
are shown in Figs.~\ref{fig_cleo_glue}(a)~and~\ref{fig_cleo_glue}(b).  
Fig.~\ref{fig_cleo_glue}(a) shows a histogram for our data, 
a curve approximating the experimental line shape for 
the not-found in our analyses $f_j(2220)$ and 
the analytical shape chosen to approximate combinatorial 
and two-photon continuum backgrounds, arrows show 
the signal region used to estimate the upper limit. 
Fig.~\ref{fig_cleo_glue}(b) shows points with the errors for our data, 
a histogram describing the $f_j(2220)$ experimental line shape 
obtained from our signal MC simulation, a curve that shows the result of 
binned maximum likelihood fit to separate the sample into signal and background 
contributions, the insert shows the signal region. 
More information on CLEO antisearches for glueballs 
can be found in publications that describe these analyses\cite{cleo_glue_1,cleo_glue_2}. 

\section{Transition Form Factors}

\begin{figure*}[t]
\centering
\includegraphics*[width=180mm,height=150mm]{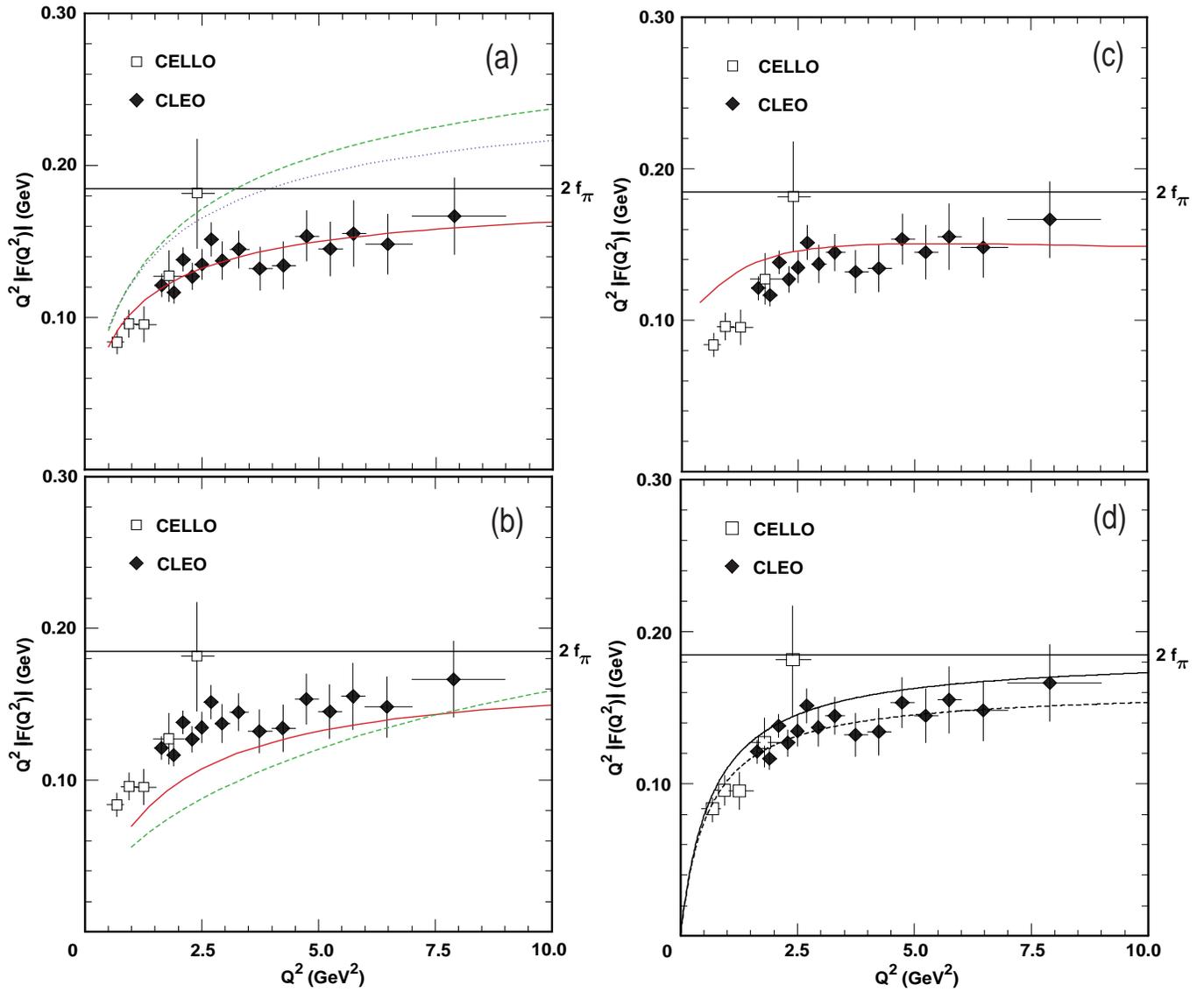}
\caption{CLEO results on $\gamma^*\gamma \to \pi^0$ production. 
See the text for more information.}
\label{fig_cleo_ff}
\end{figure*}

In 1998 we published the results of our extensive analysis\cite{cleo_ff} of the 
$\gamma^*\gamma \to {\cal R}$ transition form factors for three resonances ${\cal R}$: 
$\pi^0$, $\eta$ and $\eta^\prime$. It turned out to be an important 
publication providing data that helped, among other applications, 
to reduce theoretical uncertainties in form factors predictions 
for semileptonic and hadronic decays of $B$ and $D$ mesons. 
Fixing these form factors is necessary for 
extracting the values of CKM matrix elements 
from data collected at existing and future experimental 
facilities. Our data were also used to reduce 
the theoretical uncertainty in hadronic contribution 
from light-by-light scattering to 
the result of the Muon $g-2$ experiment. 
According to a number of theoretical papers our $\pi^0$ result 
proves the transition to perturbative QCD region at relatively low 
momentum transfer (negative squared mass of the highly off-shell photon). 
Our publication\cite{cleo_ff} also 
has references to theoretical papers 
where this conclusion has been challenged. 
We compare our $\pi^0$ result with some of available theoretical 
predictions in Fig.~\ref{fig_cleo_ff}. 
This figure also shows the results of CELLO experiment\cite{cello} at 
lower values of momentum transfer $Q^2$.
The horizontal axis is momentum transfer $Q^2$ and 
the vertical axis is the product of $Q^2$ with the 
absolute value of the $\gamma^* \gamma \to \pi^0$ transition 
form factor. Notice that this form factor is proportional 
to the square root of the observed cross section after 
effects of the $e^+e^- \to e^+e^- \pi^0$ kinematics are removed.  
Horizontal line shows the well-defined $Q^2 \to \infty$ limit of pQCD\cite{BL}. 
Fig.~\ref{fig_cleo_ff}(a) compares our results 
with pQCD-inspired prediction\cite{jakob_pi0} that uses 
(the unique) asymptotic\cite{BL} wave function 
(shown with solid curve) 
and 
Chernyak-Zhitnitsky (CZ) model\cite{cz} wave function (shown with dashed curve) 
to approximate non-perturbative effects. 
The dotted curve shows the effect of running $\alpha_s$ 
on the latter prediction. 
Fig.~\ref{fig_cleo_ff}(b) compares our results 
with another pQCD-based prediction\cite{caofang}, 
the solid curve is for asymptotic wave function 
and the dashed curve employs the CZ model distribution amplitude. 
Fig.~\ref{fig_cleo_ff}(c) compares our results 
with theoretical prediction\cite{rad} based on QCD sum rules method\cite{arkady}. 
Eventually, such methods should help to describe 
strong interactions in non-perturbative domain 
from first principles. 
Fig.~\ref{fig_cleo_ff}(d) compares our results 
with interpolation\cite{BL} suggested by Brodsky and Lepage 
(solid curve) that obeys both $Q^2 \to 0$ and $Q^2 \to \infty$ QCD limits. 
Amazingly, our results agree well with 
the $Q^2 \to \infty$ pQCD prediction 
corrected to first order in $\alpha_s$ (not shown in figure). 
Dashed curve shows the result of a phenomenology-based pole-mass fit 
to our data that does not obey $Q^2 \to \infty$ pQCD limit. 
Many other theoretical predictions are available in the literature. 
More information is available in our publication\cite{cleo_ff}. 
\newline \indent Our results for $\gamma^*\gamma \to \eta$ and 
$\gamma^*\gamma \to \eta^\prime$ transition form factors 
(plots are not shown in this review) are in full agreement 
with the prediction based on mixing measured 
from two-photon partial widths for these resonances\cite{kroll_mixing}. 
More interestingly, the $\eta^\prime$ result was utilized\cite{feldman_charm} 
to challenge the hypothesis of possible intrinsic charm\cite{ariel_charm} in 
$\eta^\prime$ suggested to explain the anomalously large 
branching fraction\cite{cleo_charm} discovered and measured by CLEO 
for the decay $B \to K \eta^\prime$. 

\section{Other Opportunities}

Existing CLEO II and II.V data could be used for a variety of 
other interesting two-photon analyses probing dynamics of 
strong interactions. These include detailed analyses 
of $K_s K \pi$, $\eta \pi\pi$, $\pi^0\pi^0$ final states 
at invariant masses below 2.5 ${\rm GeV/c^2}$ 
where glueball searches could be greatly extended, 
a study of a quantum mechanical interference between 
two-photon and bremsstrahlung production mechanisms for $\pi\pi$ pairs 
sensitive to relative strong phase between corresponding amplitudes, 
possible search for the $\eta_c^\prime$, analyses of axial-vector mesons 
produced when at least one of the photons is off-mass shell and 
many other projects. 
Possible analysis of $\pi^0$ production by two off-shell photons 
deserves special mention as this study could give a definitive 
answer about pQCD applicability at moderately high momentum transfer. 
\newline \indent The specifics of the new CLEO III data could allow us to probe the threshold 
behavior of a number of two-photon hadronic cross sections. 
The optimistic prognosis here comes from the fact that no 
filtering has been done on CLEO III to reduce beam-gas contamination, 
courtesy of powerful data acquisition system and event storing capabilities of 
the new experiment. It should be noted, however, that the new 
data samples of low final state particle multiplicities 
will not be ready for the CLEO user-level analysis for some time. 

\section{Advantages of CLEOc}

As the $B$ factories at SLAC and in Japan came on-line 
and proved to be a great success, CLEO experiment is 
changing the priorities and is about to start the new 
experimental program in the region of $e^+e^-$ center-of-masses 
energies between 3 and 5 GeV. While the range of invariant masses 
of two-photon systems accessible at this new facility 
is going to be below $\approx$1.5 GeV, 
there are certain benefits associated with reduced Lorentz boost 
for low-mass two-photon production. 
For example, our estimates show that with the same detector 
geometry, the number of detected and reconstructed $\pi^0$ 
events accompanied by detected electron or a positron 
per unit of $e^+e^-$ luminosity could be by the order 
of magnitude higher than at $\Upsilon(4S)$ energies. 
Same estimate applies to $\pi\pi$ pairs that should 
allow us to probe the threshold production 
important for chiral perturbation theory predictions. 
The increase in the number of events is achieved 
by becoming sensitive to the region 
of lower momentum transfer. Therefore, at CLEOc we lose 
sensitivity in the perturbative region of high momentum transfer 
but become able to probe highly non-perturbative 
region of low momentum transfer. 
The measurements of the $\gamma^*\gamma^* \to \pi^0$ transition form factor 
and $\gamma^* \gamma \to \pi\pi$ cross sections at threshold are 
among highlights of two-photon program at CLEOc. 
We would like to emphasize that these measurements 
are also among 
interesting opportunities potentially available 
at the PEP-N experiment. 

\section{Conclusions}

It has been known since long time ago that two-photon processes 
provide clean laboratory to study properties of strong interactions. 
The measurements of two-photon partial widths allow us to test 
the models of binding potential and mesons decays. 
When combined with results from radiative decays of 
$J/\psi$ and, in the near future, of $\Upsilon$ resonances, 
two-photon partial widths can tell us about possible mixing 
of mesons with glueballs. Extending $\gamma^*\gamma$-meson transition 
form factors measurements to the axial-vector sector 
should allow more tests of model wave functions and 
theoretical predictions eventually derived from the first principles. 
These measurements help 
to fix hadronic uncertainties in precise measurements 
of CKM matrix elements and in searches for new physics 
at existing and future experiments. 
Two-photon measurements at the $e^+e^-$ machines continue to play 
important role in learning about properties of strong interactions.  
\newline \indent The credit for the analyses summarized in this short 
review belongs to the members of the CLEO collaboration, 
past and present. These usually challenging physics analyses 
are the product of thorough studies done by many people. 
Besides efforts of my CLEO colleagues, CESR accelerator physicists 
and support personnel, I would like to acknowledge 
interesting and stimulating discussions 
with Stanley Brodsky, Thorsten Feldman, Iliya Ginzburg, 
Peter Kroll, Kirill Melnikov, Valery Serbo and Arkady Vainshtein. 
It is my pleasure to thank the organizers of the PEP-N workshop 
for creating productive and stimulating atmosphere.


\end{document}